# CALCULATION OF ACCEPTANCE OF HIGH INTENSITY SUPERCONDUCTING PROTON LINAC FOR PROJECT-X*

A. Saini[#], K. Ranjan, University of Delhi, Delhi, India
N. Solyak, S. Mishra, V. Yakovlev, FNAL, Batavia, IL 60510, U.S.A.

*Abstract*
Project-X is the proposed high intensity proton facility to be built at Fermilab, US. Its Superconducting Linac, to be used at first stage of acceleration, will be operated in continuous wave (CW) mode. The Linac is divided into three sections on the basis of operating frequencies & six sections on the basis of family of RF cavities to be used for the acceleration of beam from 2.5 MeV to 3 GeV. The transition from one section to another can limit the acceptance of the Linac if these are not matched properly. We performed a study to calculate the acceptance of the Linac in both longitudinal and transverse plane. Investigation of most sensitive area which limits longitudinal acceptance and study of influence of failure of beam line elements at critical position, on acceptance are also performed.

## INTRODUCTION

Project-X is a high intensity multi megawatt (MW) proton facility (Fig .1) to be built at Fermilab [1]. It enables a world-class Long Baseline Neutrino Experiment (LBNE) via a new beam line pointed to DUSEL in Lead, South Dakota, and a broad suite of rare decay experiments. The facility is based on 3-GeV [2], 1mA superconducting (SC) continuous wave (CW) linac. At the downstream of SC CW linac, 5-9% of the H[-] beam is accelerated in a SRF pulse linac or RCS for injection to Recycler/Main Injector and rest of H[-] beam from the linac is directed to three different experiments. Layout of Project-X facility is shown in Fig.1.

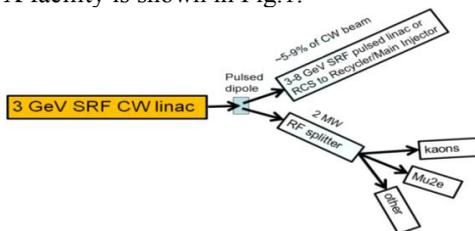

Figure 1: Project-X configuration.

The schematic of baseline configuration of the linac is shown in Fig. 2. It includes an ion source which provides 5 mA pulsed beam of H[-] ions. The beam is accelerated through the RFQ which is operated at room temperature at 325 MHz frequency. RFQ provides the beam with transverse emittance of 2.5e-7 m and longitudinal emittantce of 1.5 keV*nsec. The RFQ is followed by Medium Energy Beam Transport (MEBT) section which is used to chop the beam in order to get the time structure which is necessary to operate the different experiments simultaneously. The MEBT is followed by SC linac, which is segmented into two sections: low energy part and high energy part. The low energy section (2.5-160 MeV) uses three families of SC single spoke resonators i.e. SSR0, SSR1 & SSR2 [3] which are operated at 325 MHz. The high energy section of the SC linac (160 MeV-3.0 GeV) uses three families SC elliptical shape cavities i.e. β=0.61 & β=0.9 [4] which will be operated at 650 MHz and β=1.0 which will be operated at 1.3 GHz. Numbers of beam line elements (RF cavities, solenoids and quadrupoles) in each section along with their transition points are summarized in Table 1.

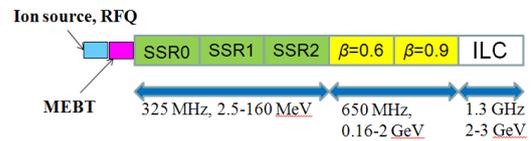

Figure. 2: Acceleration scheme for CW linac.

Table 1: No. of Elements in SC CW Linac.

|  | Cavities | Solenoids | Quads | CM |
| --- | --- | --- | --- | --- |
| SSR0 | 26 | 26 | 0 | 1 |
| SSR1 | 18 | 18 | 0 | 2 |
| SSR2 | 44 | 24 | 20 | 4 |
| β=0.6 | 42 | 0 | 21 | 7 |
| β=0.9 | 96 | 0 | 12 | 12 |
| ILC | 64 | 0 | 8 | 8 |

## GENERAL

Use of six different families of cavity result in segmentation of linac which can introduce discontinuity in focusing period. It ultimately results in shrinkage of acceptance. In this paper, we present the study of acceptance in both planes i.e. longitudinal and transverse for SC CW linac of Project-X facility. Investigation of most sensitive area which limits longitudinal acceptance and study of effects of failure of beam line elements at critical position, on acceptance are also presented. We would also like to emphasize that latest design of SC CW linac [5] differs significantly, but studies presented in this paper are used as guidelines for designing of latest lattice.

## STUDY OF ACCEPTANCE FOR SC CW LINAC

Calculations have been performed to study the longitudinal and transverse acceptance using multi particle tracking code TRACK [6]. TRACK performs calculation of acceptance for each plane (longitudinal and transverse plane) separately. It generates particle distribution with large emittance in one of plane and relatively very small emittance in other planes. Space charge effects are not considered during acceptance

*Work supported by IUSSTF and Fermi Research Alliance, LLC under Contract No. DE-AC02-07CH11359 with the US Department of Energy
[#]asaini@fnal.gov

calculation for the same reason and studies are performed for zero current.

*Calculation of Longitudinal Acceptance*

Longitudinal acceptance is calculated for ideal lattice which means misalignments of beam line elements, phase errors and failure of beam line elements are not considered. Fig. 3 shows initial coordinates of particles (yellow region) in longitudinal phase space ($\Delta W$, $\Delta t$) which are transmitted through the linac out of initial distribution (magenta region). Area defined by yellow boundary gives longitudinal acceptance of linac. Blue ring encloses area occupied by $5\sigma$ beam. The beam is laid well inside acceptance, thus no losses have been observed for ideal lattice. Particles which fall out of longitudinal acceptance i.e. yellow region in Fig. 3 are no longer accelerated in linac, hence are not matched downstream quadrupole lattice which is designed for fully accelerated particles and consequently they are lost due to transverse mismatch.

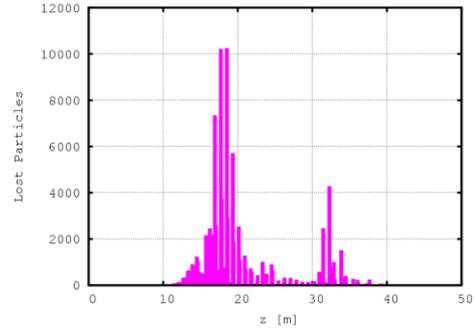

Figure: 4 Distribution of lost particles after tracking through linac.

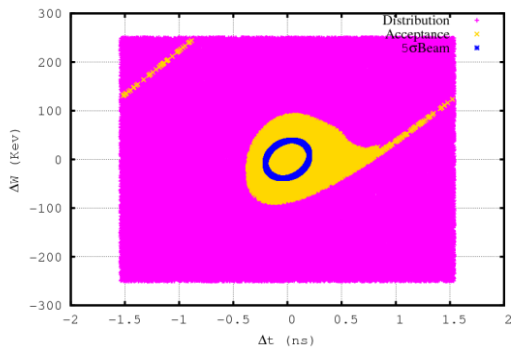

Figure.3: Longitudinal Acceptance of SC CW linac.

Linac is divided into three sections (325 MHz, 650 MHz & 1.3 GHz) on the basis of frequency. It is done in order to achieve high transverse acceptance but at the same time frequency jump may introduce a discontinuity in average longitudinal force per focusing period and limits longitudinal acceptance if transitions are not performed properly. In CW regime where tolerances are very high, thus it is necessary to achieve large acceptance through complete linac to limit beam losses and hence possible activation. In order to find most sensitive region which limit acceptance, distributions of lost particles are plotted along the length of linac (Fig. 4). It can be seen in Fig. 4 that major portion of particle losses happen at transition between SSR0-SSR1 sections. No beam losses observed after 40m tracking through linac. Thus, longitudinal acceptance is limited after SSR1 section and this acceptance is propagated through rest of linac. To study the nature of longitudinal acceptance through other sections, calculation is performed for transition between successive sections. Fig. 5 shows longitudinal acceptance for different cases of transition between neighbouring sections.

It can be easily observed in Fig. 5 that longitudinal acceptance is smallest during transition of SSR0-SSR1 section. It is required to match properly to this section to achieve larger longitudinal acceptance for beam propagation through linac.

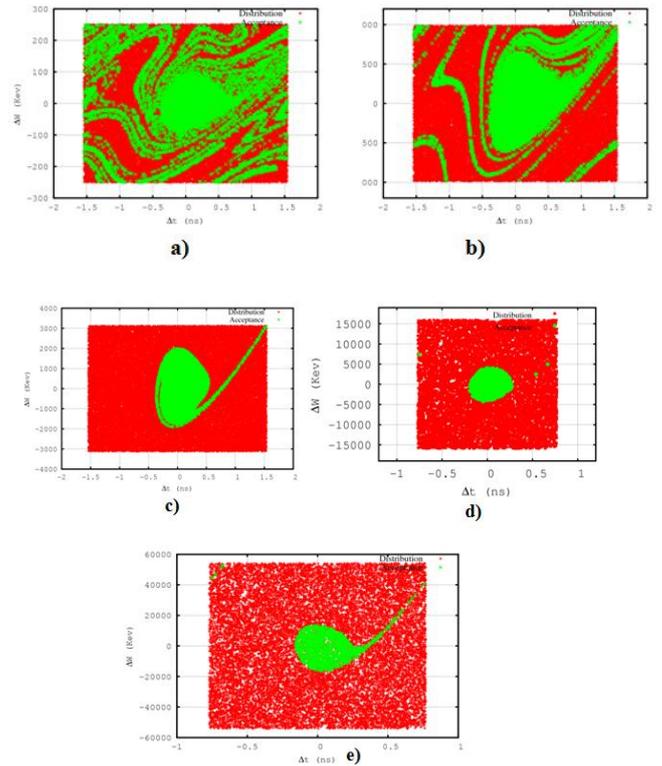

Figure: 5 Longitudinal acceptance through a) SSR0-SSR1 b) SSR1-SSR2 c) SSR2-Beta 0.6 d) Beta 0.61-Beta 0.90 e) Beta 0.90 – ILC.

*Calculation of Transverse Acceptance*

Transverse acceptance is as important as longitudinal acceptance in linac. Transverse acceptance for SC CW Project-X linac has been calculated and shown in Fig. 6

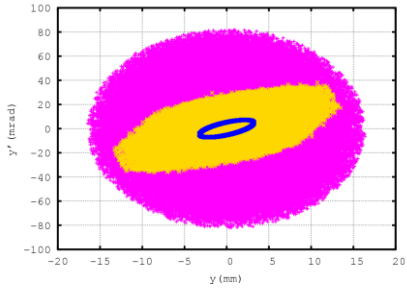

Figure 6: Transverse Acceptance through SC CW Linac.

Fig. 6 shows phase space area available for beam acceleration in vertical plane (yellow region) out of initial available phase space area (magenta region). Area under blue ring shows area occupied by beam in transverse phase space (y, y'). It is found that vertical acceptance is same as horizontal acceptance of linac, thus calculation is shown here only for vertical plane.

## CALCULATION OF ACCEPTANCE WITH FAILED ELEMENTS IN BEAMLINE

Failure of the beam transport elements like cavity, solenoid and quadrupole alters the focusing period of the beam, resulting in a mismatch of the beam with the subsequent sections. This, in turn, reduces beam acceptance. Calculations have been performed to study changes in acceptance after failure of first RF cavity in SSR0 section.

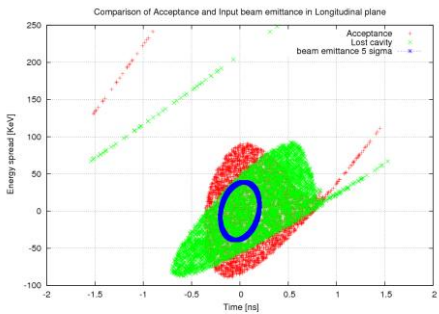

Figure 7: Acceptance through SC CW linac with failed cavity (green) & without failed cavity (red).

It can be seen in Fig. 7 that longitudinal acceptance changes significantly after failure of first cavity in SSR0 section. The beam emiitance ellipse touches boundary of acceptance. Thus, It shows that further mismatch in longitudinal plane can bring beam ellipse out of accepted region which results in beam losses.

Lattice is retuned using compensation technique [7] to mitigate the effects of failed elements in beam line. Longitudinal acceptance is calculated for retuned lattice. It can be seen in Fig. 8 that longitudinal acceptance is restored after applying compensation technique and beam is laid well inside acceptance.

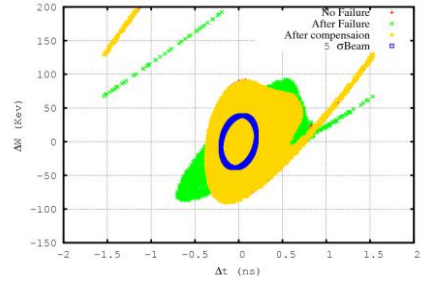

Figure 8: Longitudinal Acceptance through SC CW linac without failed cavity (red), with failed cavity (green) & after applying compensation (brown).

Study has also been preformed to investigate effects of failure of focusing elements on acceptance. Fig. 9 shows rotation of transverse acceptance after failure of first solenoid in SSR0 section. Beam ellipse (magenta) is still well inside transverse acceptance. Transverse acceptance can also be restored after applying compensation technique.

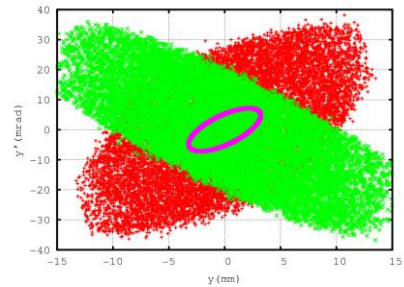

Figure 9: Transverse Acceptance through SC CW Linac with failed cavity in SSR0 section (green) & without failed cavity (red)

## CONCLUSION

Acceptance calculations have been performed for SC CW Project X linac. It is found that acceptance in longitudinal plane is limited at transition between neighbouring section. To achieve large acceptance, it is necessary to optimize matching between these sections. Longitudinal acceptance changes significantly after failure of beam line element but it is possible to restore acceptance after applying compensation echnique.